\documentclass[11pt,twoside,draft]{article}

\advance \textwidth by -5cm
\advance \textheight by -6cm
\begin{document}

\title{\bf{Filamentary, My Dear Watson!}}
%%% Author(s) and affiliation - use for single author
\def\affil#1{\begin{itemize} \item[] #1 \end{itemize}}

\author{Akhlesh Lakhtakia}
\date{}
\maketitle

\affil{CATMAS~---~Computational \& Theoretical Materials Sciences Group\\
Department of Engineering Science and Mechanics,
Pennsylvania State University\\
University Park, PA 16802--6812, USA\\
Tel: +1 814 863 4319; Fax: +1 814 863 7967;  email: AXL4@psu.edu
}

\thispagestyle{empty}

%%% Abstract
\begin{abstract}
\noindent
Developments that took place in the area of sculptured--thin--film technology
after {\em Bianisotropics '98\/} are reviewed here. Devices addressed include:
optical filters of various kinds, laser mirrors, gas--concentration sensors,
optical interconnects, interlayer dielectrics, and biosensors. Pulse bleeding
is related to the circular Bragg phenomenon
in chiral sculptured thin films. Acoustic research is also identified. 

\end{abstract}

Light from a
fluorescent street lamp filtered through the soft rain, and the leaves
of an ornamental maple tree on the other side of Baker Street
cast muted shadows on the wall opposite the bay window in the
cavernous living room of 221 B. The September night was
in full glory, as Sherlock Holmes smoked a meerschaum pipe another people's
princess had sent him as a token of her gratitude more than
a century ago.  He sat quietly, absorbed in some deep thoughts.
Only the frequency with which he released smoke rings
indicated his phrenetic turmoil.

\smallskip \begin{center} *$\,^{-}\,$*$\,^{-}\,$* \end{center}\smallskip

But, what's time in our state now? Having left aside
our mortal accouterments, Holmes and I still inhabit our 
beloved apartments in the metropolis. I continue to organize
many of the cases that Conan Doyle,
our literary agent in our former life, never published because
they were then deemed inimical to public order. Not nowadays, however.
Holmes often tells me of particularly horrid instances of murder
that the modern criminal has the means and the desire to commit
and does. That of a little girl in Boulder
and that of a big girl in Los Angeles
come to mind. Occasionally, Holmes is sufficiently intrigued to
solve particularly infamous crimes, but Conan Doyle is no longer
able to bring the criminals to book. 

The great love of Holmes these days, however, is not crime but
the science of materials. No longer able to experiment himself as he
once did on the ashes of cigars of 34 different provenances
and the herbs used by the Moluccans to shrink other people's
heads, he spends
many a night at the Imperial College library, poring over the pages of
thousands of journals and other scientific and technological
proceedings. He has an astonishing capacity to
memorize diverse facts, and he synthesizes new constructs
with uncommon fecundity. I have always felt
that the gain of Victorian crime--fighters was the loss of Victorian
natural philosophers. The recent discovery of materials with M\"obiated 
bianisotropy~---~a truly bizarre material structure that promises a thorough rewrite of
plane waves in physics textbooks~---~came of a suggestion Holmes
had whispered
in the ears of a visiting San Doggo del High physicist as he slept late
one evening, in the Imperial
College library, 
about two years ago. More often than not, he is unable to find
a competent medium for his ingenious ideas; though Holmes did
manage to masquerade as Daedalus for several years in the
pages of {\em New Scientist\/}.

I remember clearly that in 1959 he came across a report in {\em Nature\/}
by two Nottingham engineers working for a 
Swedish company.\footnote{N.O. Young and J. Kowal,
``Optically active fluorite films,"
{\em Nature\/} {\bf 183}, 104--105, 1959.} These
worthies had sought fit to outsmart nature by depositing a twisted
film of fluorite on a flat glass substrate. Their method of inducing
the twist was simplicity itself: just rotate the substrate while the
fluorite adatoms fall obliquely on it. A stationary substrate results
in columnar thin films that can act like biaxial crystals;
a rotating substrate can give rise to sculptured thin films that
are ever more complex. 

Neither one of the Nottingham duo could have verified the
true nature of their films as scanning electron
microscopes had yet not made their
{\em d\'ebut\/}, though they did have an inkling that
the microstructure was some limiting case of a \u{S}olc filter.
Holmes, who can see matters more clearly than scanning
electron microscopes because he can perform  wavelet
transforms on his nebular personal wavefunction, returned
that January dawn from the Imperial College library, declaring that he
knew the microstructure. 

``Filamentary, my dear Watson,"
he had loudly shouted for all to hear~---~but no person then alive did.
The Nottingham paper gathered dust for many decades. Only two reports were
filed over 35 years by
the Baker Street Irregulars, now no longer confined to their corporeal
sheaths but clad in black instead: 
A Scotsman living in Arizona once mentioned the paper during an
antipodean seminar, and a Louisiana physicist wrote of something similar.
And then in 1995, 
an engineer from Pencilmania  and a mathematician
from the nether Caledonia discussed the films in a Royal Society paper, not at all
aware of the Nottingham paper. Shortly thereafter, impelled by the
Pencilmaniac engineer, a Canadian undergraduate student
was able to duplicate the 1959 feat and scrutinize the arrays
of parallel straight curls that chiral sculptured thin films are.
Two years later, the Nottingham report came to light again, and
today occupies its rightful place in the scientific literature
on thin films. What an amazing confirmation of Holmes'
foresight!

\smallskip \begin{center} *$\,^{-}\,$*$\,^{-}\,$* \end{center}\smallskip

Cogitating for over an hour in complete silence, Holmes said out aloud: ``What do you 
make of the prospects
of sculptured thin films, my esteemed Watson?"

Taken aback, I mumbled: ``How the deuce did you know that?" 

``You know my methods, Watson. Apply them. After the last
gawking humans left our abode in the evening, you were nowhere
to be found. Looking at your desk, I saw the letters W I L Y  traced
in the dust on your table, and an American envelope 
with a coyote stamp lying on the floor. Clearly, our transatlantic friend John
had communicated with you through an infernal countryman
of his who spent almost the whole day here in 221 B. I called one
of my men in black,
who confirmed that you had been seen crossing the Brooklyn Bridge.
On your return, I detected perfume clinging to you. You must
have been looking over the shoulder of a copy--editor in New York."

``Yes, but how did you know that John's missive concerned
sculptured thin films?" 

``Obsessed with 
these films you have been for some months, haven't you?"

``Yes. It does sound so simple, Holmes, when you explain it.
Anyhow, there is this new book being published,\footnote{O.N. Singh and
A. Lakhtakia (eds),
{\em Electromagnetic Fields in Unconventional Materials
and Structures\/} (Wiley, New York, 2000).} 
and I sneaked
a preview of Chapter 5,
just to apprise you of the latest developments."

``Pray do," quoth he, as he slowly floated up to the top of the ottoman.

I began:
``Well, it seems that matters have advanced quite considerably during the
past two years. You may recall, Holmes, the addresses that
the Pencilmaniac engineer
had delivered at {\em Bianisotropics'97} in Glasgow and {\em Bianisotropics '98}
in Braunschweig. In his first address, he was quite speculative about
sculptured thin films. Reliable methods of producing them were 
still in their infancy; and the optical rotation spectrum of only one chiral
sculptured thin film had been measured by then~---~and that too was 
crude and incomplete. The Pencilmaniac engineer's group took many theoretical
strides following the Glaswegian meeting, which were duly reported
in Germany. The theoretical basis of optical devices~---~such as circular polarization
filters, laser mirrors, and gas--concentration sensors~---~was also
firmed up between those two meetings. 

``Several developments took place after {\em Bianisotropics '98}, many
of which are reported in Chapter 5 of this new book. Optical interconnects
as well as interlayer dielectrics are among the emerging applications
of the sculptured--thin--film nano\-technology. Some acoustics research
has also taken place, but it is definitely in the theoretical stage only.
Most importantly,
at least three types of optical devices have been made with the new
serial bideposition technique
in New Zealand, in collaboration
with the Pencilmania group.

``The first optical device is a 
circular polarization filter that allows either left-- or right--handed
plane waves to pass through, but not both. Of course, on enquiry both Williams
bragged that they have known from Wilhelm's
time that such filters can work only in
relatively narrow wavelength--regimes.

``The next device combines the filtering action with polarization--inversion.
The circular polarization filter is capped by a columnar thin film 
which functions as a half--wave plate at a certain wavelength within the
Bragg regime. The device thus is really the first two--section sculptured
thin film ever made, and fully justifies your confidence in the
concept of sculptured thin films.

``Finally, the third device is a spectral hole filter. While a chiral sculptured
thin film
is being deposited, the substrate rotation is temporarily stopped and
then resumed after a while. This results in the production of a reflection hole
that punctures the Bragg regime, when all parameters are properly chosen.
The bandwidth of the hole is 11~nm, which is comparable to the
10~nm holes produced commercially by holographic techniques."

Holmes interjected: ``That is a three--section sculptured thin film then, isn't it?
But they have done even better with reflection holes. Instead of stopping the
rotation of the substrate, they now simply give a quick orthogonal twist.
They have achieved the same type of hole with a two--section sculptured thin film."

``How do you know about that?" I asked, to which Holmes replied that he
had sneaked into an editorial office at a London physics department.
He was also tickled pink to find that an Imperial College lecturer had joined
the Pencilmania--NZ collaboration.

Holmes continued: ``I can see the possibilities of 
highly sensitive gas--concentration sensors
in those spectral holes. Did you find any evidence in the
book chapter?" My reply in the
negative made him pensive for a while. ``Anything else in that book?" he questioned,
and I mentioned that the Pencilmania group had undertaken the incorporation
of further verisimilitude in their research by assuming Lorentzian
dependences for the constitutive parameters. ``Good," he went on, ``Hendrik
will be pleased. But we must find more on what's afoot in Pencilmania."

Dawn was about to break as he uttered those words. Soon the fog would roll
out of Heathrow and aeroplanes would begin landing there, bringing another
clutch of Holmesians to 221 B. It was time to retire for the day, but Holmes
went out to speak to one of his men in black.

\smallskip \begin{center} *$\,^{-}\,$*$\,^{-}\,$* \end{center}\smallskip

The great merit of disembodiment is that the Baker Street Irregulars
 can travel quite fast. While
speeds close to or more than a tenth of the speed of light are not advised,
lest a blue glow be emitted, during the last few
years the men in black have been able
to go to almost anywhere in the world with the help of the Internet. 
Constantly 
bumping into copper atoms used to be a hazard, but optical fibers now
provide them with a smooth Alpine slide, even to Papua Niugini.
Holmes and I also use the same mode of travel, incidentally.

\smallskip \begin{center} *$\,^{-}\,$*$\,^{-}\,$* \end{center}\smallskip

As the shadows filtered in again, ushering a new twilight, Holmes
began to receive reports from the Baker Street Irregulars who had
combed through the editorial offices of a multitude
of physics and engineering
journals. That the Lorentzian route had been taken by the Pencilmaniac
was confirmed not only from France but also from Germany and Italy. No less
than three editorial outfits from the three countries had
become the Pencilmaniacs' accomplices, so to say. Sound research
had been conducted and reported.

An interesting development was the modeling of individual columns
in chiral sculptured thin films as ensembles of
inclusions laid out end--to--end, not unlike strings of
sausages. Spectral maximums of various observable properties
had been examined 
 as functions of inclusion shape, volume fraction
and orientation~---~for eventual use in
computer--aided  design.

Even more astonishingly,
research had spilled over from the frequency into the time domain.
After solving the Maxwell equations directly in the time domain,
the spatio--temporal anatomy of the
circular Bragg phenomenon exhibited by chiral sculptured thin films
had been bared, much to the delight of the doctor in me. Pulse bleeding
had been shown to occur in the Bragg regime under certain circumstances.
Holmes rubbed his hands with glee, as he wondered about the
encounter of femtosecond pulses with chiral sculptured thin films,
and cautioned me to be careful: if these new materials were to
be used for wavelength division multiplexing and demultiplexing,
we would have to choose suitable polarizations for future Internet
travel. ``Better pack a few extra polarizations, Joseph, just to stand
out from the hoi polloi."
In some of his lighter moments, Holmes fancies himself as
Sir Andrew Lloyd Webber giving stage directions!
 
``Watson, the Pencilmaniac engineer is moving just
too fast. I fear I am unable to delve into his brain, because
he never puts on the virtual reality headset I had surreptitiously
suggested the head of his department to provide him with. But his
thoughts cannot elude detection. Unable to remember
anything for too long, he commits all his ideas to writing.
Somewhere in his office, a blueprint of his plans must be
hidden; and I must lay a trap for him." Holmes pronounced each word
with deliberation, in his usual calm manner. Every comma
and every semi--colon, not to mention full stops, were marked
by pauses of the right duration. I could almost
feel that gears were whirring and lights were flashing inside that
powerful intellect of his. And then he went out $............$ with one
of his men in black.

\smallskip \begin{center} *$\,^{-}\,$*$\,^{-}\,$* \end{center}\smallskip

Holmes did not come back in the morning, and was
away the next night as well. Late in the following afternoon,
as I reposed on a horsehair sofa in the attic away from the prying
throngs of visitors, I became aware of a tall cylindrical
object with a faint green glow making its way towards me.
If I had a skin, I would have jumped out of it. Golly, whatever
could that dreadful apparition be? I froze in terror, as peals
of laughter rang out. 

``What's up, Doc?" a guttural {\em patois\/}
issued  from that object. But Holmes couldn't fool me. He 
is certainly a master of mimicry, but long association with him
helped me see right through him. 

``What are you disguised as?"

``What else? A Pencil, of course."

``Aha! A pencil for the Pencilmaniac!!"

``Exactly! Let's be off to Pencilmania, where the day is 
just about middle--aged now. Our quarry must be working in his office,
where we must corner him."

Hitching a ride on the telephone cable, we exited 221 B. Milliseconds
later, we
swung over to a British Telecom fiber. A skip over the splicing with
an MCI fiber, and we negotiated the Atlantic {\em via\/} a satellite
faster than you can say `WorldCom.' 

As we were landing on the desk of the
Pencilmaniac engineer through his computer, he began to raise his
right hand from his lap. He opened the side--drawer. Nimble as
a humming bird, Holmes slid into the pencil case 
lying inside the drawer. A piece of
paper lay on the desk, covered with chemical symbols such
as [Ru(2,2$'$--bipyridine)$_3$]Cl$_2$ and
[Ru(1,10--phenantholine)$_3$]Cl$_2$, and with the letters
A, T, C, G and U strewn all over.  A firefly had been doodled in a corner,
with the words {\em Lucifer~---~Son of Morning\/} written below
in a cursive hand.

The Pencilmaniac's left hand took out a pencil and drew several parallel lines
on another sheet of paper. Those lines glowed green! Holmes had
been able to get inside the engineer's brain! The Pencilmaniac
continued to make a schematic, occasionally labeling certain layers.
I did not understand the diagram, but Holmes did.

\smallskip \begin{center} *$\,^{-}\,$*$\,^{-}\,$* \end{center}\smallskip

An hour later, after the Pencilmaniac engineer had closed shop
for the day, and his office was bathed in vivacious darkness,
Holmes emerged from the pencil case. Triumphantly, he declared,
``I just want to say one word to you $..........$ Biosensors."

\bigskip
\noindent {\it Author's note: If you enjoyed the story, 
please write or call for the following publications:}

\def\refname{\normalsize {\bf Electromagnetic References}}

%%% References

%\def\refname{\normalsize \centerline{\bf Acoustic References}}
\def\refname{\normalsize {\bf Acoustic References}}

%%% References

\end{document}